\documentclass[10pt,letterpaper]{article} 
\usepackage[english]{babel}
\begin{document} 
\newcommand{\pl}{\partial}
\newcommand{\beq}{\begin{equation}}
\newcommand{\eeq}{\end{equation}}
\newcommand{\lbl}{\label}
\newcommand{\beqnar}{\begin{eqnarray}}
\newcommand{\eeqnar}{\end{eqnarray}}
\newcommand{\beqnars}{\begin{eqnarray*}}
\newcommand{\eeqnars}{\end{eqnarray*}}
\newcommand{\ip}[1]{\mbox{(#1)} \displaystyle \qq}  
\newcommand{\dg}{^\circ}  
\newcommand{\arrow}{\rightarrow}  
\newcommand{\goesto}{\rightarrow}
\newcommand{\s}{\\[1ex]}
\newcommand{\re}[1]{(\ref{#1})}
\newcommand{\q}{\quad} 
\newcommand{\non}{\nonumber} 
\newcommand{\bigoh}{$O$} 
\newcommand{\tnr}{\otimes} 
\newcommand{\tr}{\mbox{tr }} 
\newcommand{\trE}{\mbox{tr}_E\ }
\newcommand{\avec}{\vec{a}}
\newcommand{\bvec}{\vec{b}}
\newcommand{\uvec}{\vec{u}}
\newcommand{\vvec}{\vec{v}} 
\newcommand{\pvec}{\vec{p}}
\newcommand{\qvec}{\vec{q}}
\begin{center}
\large
Comments of Stephen Parrott%
\footnote{
Current contact information can be found on my web page, 
www.math.umb.edu/$\sim$sp.  
}
 concerning\\
\ \\
``An experimental test of non-local realism''\\
by\\
Simon Gr\"{o}blacher, Tomasz Paterek, Rainer Kaltenbaek, \v{C}aslav Brukner,
Marek \.{Z}ukowski, Markus Aspelmeyer, and Anton Zeilinger\\
Nature {\bf 446}, 871-875 (2007) 
\\ 
\end{center} 
\section{Introduction}

I found this paper very interesting, in fact so interesting that
I was motivated to think carefully about its assumptions and 
to check some tedious mathematics.  
I noticed what looked like a serious error in the paper's 
proof of its key inequality (9).%
\footnote{The authors have since sent me a revised proof
avoiding the error.}
In searching for an alternative
to (9), I found a simple, straightforward
proof of this inequality (based on ideas in the paper, which in turn
is based on \cite{leggett}).  This is presented in Sections 3 and 4. 

The paper seems fairly clearly written, but since it is not 
completely explicit (e.g., there are symbols whose meaning the 
reader has to guess), I was worried that I might have  
misinterpreted something.  To reduce this possibility,
the following explains my interpretation of its content in greater 
detail than usual. 

I thank the authors for their comments
and for pointing out a slip, which I have corrected.  
Of course, I take responsibility for any further errors. 

I do assume that the reader is somewhat familiar with the paper
and has it at hand.
The notation follows the paper as much as possible.  
Any undefined symbols are as in the paper.
Page numbers refer to the version www.arXiv.org/quant-ph/0704.2529v1.
I have not seen the published version, but since the arXiv version
is dated April 19, 2007 and the published version appeared days later,
I assume that they are identical, or nearly so. 
\section{My interpretation of the paper's setup} 

For ease of language when introducing the definitions, 
it will be sometimes be convenient to pretend that
probability distributions arising are discrete.
For example, the paper considers pairs of photons with polarizations
$\vec{u}, \vec{v}$, ocurring with probability density $F(\vec{u},\vec{v})$.  
I will sometimes refer to $F(\uvec, \vvec)$ as the probability 
that photon 1 has polarization  $\uvec$ and photon 2 has polarization
$\vvec$,  which would be correct language if $F$ were a discrete
probability distribution.

A source emits pairs of photons in different directions, as
depicted in Figure 2 of the paper.  
One photon goes to Alice, and the other to Bob.
 
The probability that Alice's photon has polarization  $\uvec$ and 
Bob's has polarization $\vvec$ is denoted $F(\uvec, \vvec)$.
Here $\uvec, \vvec$ represent points on the unit sphere in three-dimensional
space $R^3$.  The standard angular polar coordinates of a vector like $\uvec$
are denoted $\theta_{\uvec}, \phi_{\uvec}$, so that 
$\uvec = (\cos \phi \sin \theta, \sin \phi \sin \theta, \cos \theta)$.
This corresponds to a photon represented quantum-mechanically
by the ray
in the two-dimensional complex Hilbert space $C^2$. 
represented by the vector
$
\left[
\begin{array}{l}
\cos \theta/2 \\
e^{i\phi} \sin \theta/2
\end{array}
\right].
$

The paper considers a ``hidden variable'' $\lambda$ associated with the 
source.  Presumably, this can be thought of as a classical label attached by 
the source to each of the pair of emitted photons.
The same label is attached to each of the photons in an emitted pair,
but the label can vary from pair to pair.  

My first impression was that the authors were thinking of the source
as emitting two photons with polarizations $\uvec, \vvec$
with an additional label $\lambda$ attached to each photon, as in their
Appendix I example of an explicit non-local hidden-variable model.
(The set of possible labels $\lambda$ is allowed to depend
on $\uvec$ and $\vvec$, as in the example.)
However, this seems inconsistent with some of their later notation,
so I eventually settled on the the interpretation to be described below.
The two interpretations are essentially equivalent
(modulo technicalities), so the choice of either is a matter
of taste and notation.  

The nature of the label $\lambda$ is not specified
and is irrelevant to the proofs. 
It could be a real number in a certain range 
(depending on $\uvec$ and $\vvec$),
as in the Appendix I example, or something more complicated.

We could use a new label $\lambda^\prime$ defined as a triple 
$\lambda^\prime := (\lambda, \uvec, \vvec)$, where $\lambda$
is the ``old label'' in the viewpoint above.    
This is conceptually simpler in that there is now only one label
$\lambda^\prime$ rather than three.  In order to stay close to 
the paper's notation, from now on we write $\lambda$ instead
of $\lambda^\prime$ and work with only one label. 

The polarization
$\uvec$ of the photon received by Alice is assumed to be a function
$\uvec = \alpha(\lambda)$ of the hidden variable label attached to 
her photon, and similarly the polarization of Bob's photon
is $\vvec = \beta(\lambda)$ .%
The functions $\alpha(\cdot), \beta({\cdot})$
(which are not part of the paper's notation) are introduced 
for later convenience 
instead of writing $\uvec(\lambda), \vvec({\lambda})$;
certain distinctions are hard to make in the latter notation.

This could give a classical
explanation for correlations between the polarizations of 
Alice's and Bob's photons.  The paper's aim is to show that
such a classical explanation of observed correlations contradicts
both quantum mechanics and experiment. 

The set of possible labels is a probability space, 
whose probability measure will not be named.  
Since Alice's polarization is
a function $\uvec = \alpha(\lambda)$ of the hidden variable $\lambda$,
this induces a probability distribution $F(\uvec,\vvec)$
on the set of
polarization pairs $\uvec, \vvec$ as follows.   When the set of $\lambda$
is discrete, the probability $F(\uvec,\vvec)$ 
of a particular polarization pair
$\uvec, \vvec$ is the probability of the set of all $\lambda$
such that $\alpha({\lambda}) = \uvec$ and $\beta({\lambda}) = \vvec$. 

When $\lambda$ is a continuous variable, the mathematical object
corresponding to $F(\uvec,\vvec)$ is 
a probability measure which might be denoted 
$F(\uvec, \vvec)\, d\uvec d\vvec$ in the special case in which it is 
given by a probability density function, where $d\uvec$ and $d\vvec$
represent Lebesgue measure on the unit sphere.  We follow the paper
by using the notation of a probability density function,
with the understanding that the measure might have a singular 
part (e.g., concentrated at a point or on a line). A precise mathematical 
definition might be cumbersome, 
but the discrete case above gives the idea. 

The paper defines ``Malus' law'' as ``the well-known cosine dependence
of the intensity of a polarized beam after an ideal polarizer''.
I take this to mean the following.  Alice has an instrument to
measure polarization in any chosen direction $\avec$.  
The only possible results of the measurement are $\pm 1$.  
A reading of $+1$'' means that the observed polarization
was in the direction $\avec$ and $-1$ means that it was 
in the opposite direction $-\avec$. 
If she receives many photons with 
polarization $\uvec$, then the average reading  
is $\avec\cdot \uvec$ (which is the cosine
of the angle between $\avec$ and $\uvec$).

The paper introduces a symbol $\rho_{\uvec, \vvec}$,
giving only the cryptic explanation: ``Each emitted pair is fully defined
by the subensemble distribution $\rho_{\uvec, \vvec} (\lambda)$.''
I take this to mean that $\rho_{\uvec, \vvec}(\cdot)$ 
is a conditional probability density function:
in the discrete case, $\rho_{\uvec , \vvec}(\lambda)$
is the probability of $\lambda$ given that the polarizations
of the emitted pair was $\uvec, \vvec$.  
A precise mathematical
definition in the generality considered by the paper might be cumbersome,
but the idea is clear in the discrete case:  Given a particular
$\uvec, \vvec$ and $\lambda_0$ with $\alpha(\lambda_0) = \uvec$
and $\beta(\lambda_0) = \vvec$ , $\rho_{\uvec, \vvec} (\lambda_0)$
is defined as the probability of $\lambda_0$ divided by the probability
of the set of all $\lambda$ such that $\alpha(\lambda)= \uvec$
and $\beta(\lambda) = \vvec$.

Suppose Alice sets her instrument to measure polarization in the
$\avec$ direction, Bob sets his to measure in the $\bvec$ direction,
and the hidden variable attached to each of their photons is $\lambda$.
The paper denotes the outcome of Alice's measurement (either $+1$ 
or $-1$) as $A(a, b, \lambda)$ and Bob's as $B(a, b, \lambda)$.
The assumption that Malus' law holds is then given by the paper's
equations (1) and (2):
\begin{eqnarray}
\lbl{eq1}
\bar{A} (\uvec) &:=& \int d\lambda\, \rho_{\uvec, \vvec}(\lambda)
A(\avec,\bvec, \lambda) = \uvec \cdot \avec \q, \\ 
\bar{B} (\vvec) &:=& \int d\lambda\, \rho_{\uvec, \vvec}(\lambda)
\lbl{eq2}
B(\avec,\bvec, \lambda) = \vvec \cdot \bvec \q. 
\end{eqnarray}
(I changed the paper's first ``='' to the definition symbol 
``:='' because I think it is helpful to the reader to explicitly
distinguish between equality by definition and assertions of 
equality between separately defined quantities.) 

These equations seem  sensible in terms of the interpretation
just described in which the source emits two particles, each with
just one label (the same label) $\lambda$, which implicitly contains
the polarization information.
If one is thinking of emission of two polarizations $\uvec, \vvec$
along with an additional label $\lambda$, then in equation \re{eq1},
$A(\avec, \bvec, \lambda)$ should be written
$A(\avec, \bvec, \uvec, \vvec, \lambda) $ (or, less generally,
$A(\avec, \bvec, \uvec, \lambda) $).  In more physical language,
what Alice measures is expected to depend explicitly on the polarization
of the photon she receives.  Indeed, 
the Appendix I example writes $A = A(\avec, \bvec, \uvec, \lambda) $.  

The interpretation above (with just one label $\lambda$ which
contains the polarization information) was developed to make
sense of equations \re{eq1} and \re{eq2}.  But the two interpretations
are equivalent, modulo technicalities and notation.

\section{Why the hidden variable theory cannot reproduce quantum
mechanics}

We are interested in the following two questions.
\begin{enumerate}
\item
Can the hidden variable theory described in the previous section
reproduce the results of quantum mechanics?
\item
If not, how can we experimentally distinguish between quantum
mechanics and the hidden variable theory?
\end{enumerate}

This section presents a simple proof that the hidden variable theory cannot
reproduce the results of quantum mechanics. 
This conclusion will also follow from the results of the next section,
which answers question 2,
but we present it separately because is is a little easier 
and the result is simpler than the paper's (9).  
The proof of the next section is not much longer than
the proof of this section, but it seems less motivated.  
The present section provides the motivation, notational preliminaries,
and a few simple calculations which enter into the proof.  

Before starting, I should acknowledge that the proof's ideas 
are mostly contained in the paper under discussion, which is
based on \cite{leggett}.  Although in retrospect, the proof
seems simple, I think it would have taken me a long time to find 
it had I been given the problem without the solution hints
contained in these two references.  Any mathematician knows
that the first proof is always the hardest to construct, 
and in retrospect is often unnecessarily complicated.   

For given vectors $\avec, \bvec$, define a ``correlation function''
$C(\avec,\bvec)$ by 
\beq
\lbl{eq30}
C(\avec, \bvec) := \int d\uvec \, d\vvec\, d\lambda\,
\rho_{\uvec, \vvec}(\lambda) F(\uvec, \vvec) 
A(\avec, \bvec, \lambda) B(\avec, \bvec, \lambda)
\q.
\eeq
Here $\rho_{\uvec,\vvec}(\lambda), F(\uvec, \vvec), A(\avec, \bvec, \lambda),$
and $ B(\avec, \bvec, \lambda)$ are as defined in the paper and in the
first section above, and $\int d\uvec$ represents the integral over the unit
sphere in three-dimensional real Euclidean space (similarly for 
$\int d \vvec$).

The correlation $C(\avec,\bvec)$ is called $\langle AB\rangle$
in the paper (its equation (4)); we introduce the new notation
because we shall need to display the dependence of $\langle AB \rangle$
on the ``setting vectors'' $\avec$ and $\bvec$.

Let $\alpha := \cos^{-1} \avec \cdot \bvec$ be the angle between
$\avec$ and $\bvec$.  For a system in the singlet state (the case
considered by the paper), quantum mechanics predicts that 
$C(\avec, \bvec) = -\avec \cdot \bvec$.
In the following, it will be helpful to think of $\alpha$ as 
an acute angle (though the proof does not assume this), 
so that it is expected that $C(\avec, \bvec) \leq 0$.  
For this case, it is a little easier to work with $- C(\avec, \bvec) \geq 0$.

The paper (following \cite{leggett})  shows that: 
\beq
\lbl{eq40}
-1 + \int d\uvec\, d\vvec\, F(\uvec, \vvec) |\avec \cdot \uvec - 
\bvec \cdot \vvec | 
\leq
-C(\avec, \bvec) \leq 1 - \int d\uvec \, d\vvec \, F(\uvec, \vvec)
|\avec \cdot \uvec +  \bvec \cdot \vvec|
.
\eeq
Only the right-hand inequality will be used below, which will essentially 
result in establishing half of the paper's inequality (9).
The other half follows similarly from
the left inequality in \re{eq40},
as will be indicated in the next section.

According to quantum mechanics, for all $\avec$, 
\beq 
\lbl{eq50}
1 = - C(\avec, \avec) \leq 1 - \int d\uvec \, d\vvec\,
 F(\uvec, \vvec) |\avec \cdot (\uvec + \vvec) |
\q,
\eeq
so the integral on the right must vanish.  
Since the integrand is non-negative,
this implies that $F(\uvec, \vvec)$ must be concentrated on
the singular set of all $\uvec, \vvec$ such that
$\vvec = - \uvec$.  Restricting to this set, 
the probability distribution can be symbolically 
represented by a probability density function of just one sphere
variable $\uvec$. We denote this new probability density 
function as $F_s(\uvec)$ and rewrite inequality \re{eq40} as:
\beq
\lbl{eq55}
-C(\avec, \bvec) \leq 1 - \int d\uvec \,  F_s(\uvec)
|(\avec - \bvec) \cdot \uvec)|
\q.
\eeq

Suppose temporarily that unit vectors $\bvec \neq \pm \avec$, so that $\avec$
and $\bvec$ are contained in a unique plane. 
Following the paper and \cite{leggett}, we obtain more tractable
inequalities by averaging $C(\avec, \bvec)$ over rotations in the plane 
determined by $\avec, \bvec$ 
(i.e., rotations about the $\avec \times \bvec$ axis).
The result, which depends only on the plane of rotation and 
the angle $\alpha := \cos^{-1} (\avec \cdot \bvec)$, will be denoted 
$E(\alpha)$.  More explicitly, if $R(\sigma)$ denotes a rotation
through the angle $\sigma$ about the axis $\avec \times \bvec$, then
\beq  
\lbl{eq200}
E(\alpha) := \frac{1}{2\pi} \int d\sigma \, 
C(R(\sigma)\avec, R (\sigma) (\bvec))  
\q.
\eeq 
In this notation, $E(\alpha)$ implicitly depends on the plane of 
$\avec$ and $\bvec$. When we want to include in the notation that 
this plane is the $x$-$y$ plane,
we write $E_{xy}(\alpha)$ instead of $E(\alpha)$, and similarly
$E_{xz} (\alpha)$ denotes $E(\alpha)$ when $\avec$ and $\bvec$
lie in the $x$-$z$ plane.  

Next we derive (following the paper and \cite{leggett}) an inequality
for $E_{xy}(\alpha)$.
For any vector $\uvec = (u_x, u_y, u_z)$
on the unit sphere,
write $\uvec_{xy} := (u_x, u_y, 0) $ to denote the 
projection of $\uvec$ to the $x$-$y$-plane. Then for any vector $\qvec$
in the $x$-$y$ plane,
$\qvec \cdot \uvec = \qvec \cdot \uvec_{xy} = |\qvec||\uvec_{xy} 
| \cos \beta$,
where $\beta$ is the angle between $\qvec$ and $\uvec_{xy}$.  Hence
for $\avec, \bvec$ in the $x$-$y$ plane,
the average of $|(\avec - \bvec) \cdot \uvec |$ over rotations in that plane
is  
\begin{eqnarray}
\lbl{eq205} 
\frac{1}{2\pi} \int_0^{2\pi} d\sigma \, |(R_\sigma (\avec - \bvec)\cdot \uvec| 
&=&
|\avec - \bvec | 
 |\uvec_{xy} | 
\frac{1}{2\pi}
 \int_0^{2\pi} d\tau \, |\cos \tau| \nonumber\\
&=& \frac{2}{\pi} |\avec - \bvec| |\uvec_{xy}|\q, 
\end{eqnarray}
where the integration variable was changed from $\sigma$ to
$\tau := \beta - \sigma$, with $\beta $ the angle between 
$\uvec$ and $\avec - \bvec$. 
Combining this with inequality \re{eq55} gives 
\begin{eqnarray}
\lbl{eq210}
-E_{xy}(\alpha) &\leq& 1 - \frac{2}{\pi} |\avec - \bvec|
\int d\uvec\, F_s(\uvec) | \uvec_{xy}| \\
&=& 1- \frac{4}{\pi} |\sin \frac{\alpha}{2} | 
\int d\uvec\, F_s(\uvec) | \uvec_{xy}| \q,
\end{eqnarray}
where the last line follows from the routine calculation
$$
|\avec - \bvec|^2 = 2 - 2\avec \cdot \bvec = 2(1 - \cos \alpha)
= 4 \sin^2 \frac{\alpha}{2} .
$$ 

It is hard to deduce more from inequality \re{eq210} 
without specific knowledge of the probability density $F_s (\uvec)$.
But adding the $x$-$y$ and $x$-$z$ versions of \re{eq210} 
gives something useful:  
\beq
\lbl{eq220}
-E_{xy}(\alpha) - E_{xz}(\alpha) \leq   
2- \frac{4}{\pi} |\sin \frac{\alpha}{2} | 
\q.
\eeq
Here we have used the facts that 
$\int F_s ( \uvec) d\uvec = 1$ and that 
$|\uvec_{xy}| + |\uvec_{xz}| \geq 1$.  (Proof: 
$(|\uvec_{xy}| + |\uvec_{xz}|)^2 \geq  |\uvec_{xy}|^2 
+ |\uvec_{xz}|^2 = u^2_x + u^2_y + u^2_x + u^2_z \geq u^2 = 1$.)

The argument just given assumed that $C(\avec, \avec) = -1$,
which implies that $F(\uvec , \vvec)$ is concentrated
on $\vvec = - \uvec$. 
If $F(\uvec,\vvec)$ is not concentrated on $\vvec = - \uvec$, then 
$E_{xy}(0)$
gives some information about $F(\uvec, \vvec)$ for $\vvec \neq \uvec$.
This suggests that it might be productive to look at 
$$
-E_{xy}(\alpha) - E_{xy} (0) \q,
$$
as the paper does.

\section{Testing the hidden-variable theory}
Finally, we give a proof of 
the paper's (9) without assuming that $C(\avec, \avec) = -1$. 
We use the notation of the last section, along with some 
simple facts established there.

Apply inequality \re{eq40} to obtain
\begin{eqnarray}
\lbl{eq300}
-C(\avec, \bvec) - C(\avec, \avec) &\leq&  
2 - \int d\uvec\, d\vvec\, F(u,v)
[|\avec \cdot \uvec + \bvec \cdot \vvec| + 
| \avec \cdot \uvec + \avec \cdot \vvec|] \nonumber\\
&=&2 - \int F(\uvec, \vvec)
[|\avec \cdot \uvec + \bvec \cdot \vvec| + 
|-\avec \cdot \uvec - \avec \cdot \vvec| \nonumber\\
&\leq& 2 - \int F(\uvec, \vvec) | (\bvec - \avec) \cdot \vvec|] 
\q,
\end{eqnarray}
where the last line comes from the triangle inequality, 
$|\pvec| + |\qvec| \geq |\pvec + \qvec|$.

Let $\alpha := \cos^{-1} \avec\cdot\bvec$ be the angle between $\avec$
and $\bvec$. Average over rotations in the $x$-$y$plane to obtain 
\begin{eqnarray}
\lbl{eq310}
- E_{xy}(\alpha)  - E_{xy}(0)
&\leq & 2 - |\bvec - \avec| \frac{2}{\pi} 
\int d\uvec\, d\vvec\,  F(\uvec, \vvec)|\vvec_{xy}| 
\nonumber
\\
&=& 2 - \frac{4}{\pi} |\sin \frac{\alpha}{2} | 
\int d\uvec\, d\vvec\,  F(\uvec, \vvec)|\vvec_{xy}| 
\q.
\end{eqnarray}
The same procedure using the 
left inequality in \re{eq40} yields
\begin{eqnarray*}
C(\avec, \bvec) +  C(\avec, \avec) &\leq&  
2 - \int d\uvec\, d\vvec\, F(u,v)
[|\avec \cdot \uvec - \bvec \cdot \vvec| + 
| \avec \cdot \uvec - \avec \cdot \vvec|] \\ 
&\leq& 2 - \int F(\uvec, \vvec) | (\bvec - \avec) \cdot \vvec|] 
\q,
\end{eqnarray*}
so
$$ 
E_{xy}(\alpha) +  E_{xy}(0) \leq 
2 - \frac{4}{\pi} |\sin \frac{\alpha}{2} | 
\int d\uvec\, d\vvec\,  F(\uvec, \vvec)|\vvec_{xy}| 
\q.
$$
Combining this with \re{eq310} gives
\beq
\lbl{eq315}
|E_{xy}(\alpha) + E_{xy}(0)|
\leq 
2 - \frac{4}{\pi} |\sin \frac{\alpha}{2} | 
\int d\uvec\, d\vvec\,  F(\uvec, \vvec)|\vvec_{xy}| 
\q.
\eeq
Do the same for the $x$-$z$ plane and add the results, recalling from 
the last section that $|\vvec_{xy}| + |\vvec_{xz}| \geq 1$,  
to obtain the paper's (9): 
$$
| E_{xy}(\alpha)  + E_{xy}(0)| + |E_{xz}(\alpha) + E_{xz}(0)| 
\leq 4 - \frac{4}{\pi} |\sin \frac{\alpha}{2} |
\q.
$$
for the particular choice of orthogonal planes $x$-$y$ and $x$-$z$.  

Of course, the proof just given applies to any two orthogonal
planes---the particular choice of planes was made to simplify 
the notation.  The paper's statement of its (9) appears to
apply to {\em any} two planes, not necessarily orthogonal.
However, its proof does explicitly assume orthogonal planes
(on the top of its page 13), so I assume this was intended.  
\section{Statistical methods} 

The paper does not completely explain its statistical methods,
and I'm not sure I can agree with what is explained.  
I have questions about the standard deviations claimed.
The paper states that ``the errors [presumably meaning standard
deviations] are calculated assuming that the counts follow a 
poissonian distribution''.  I don't understand this assumption.
I'm not sure precisely what it means, and under all interpretations
which have occurred to me, it seems questionable.

If we were measuring the number of counts observed by Alice in
a given time interval (say the 10 sec. mentioned on p.\ 5, during
which Alice observes about 95,000 counts), {\em that} would
be expected to follow a Poisson distribution:%
\footnote{The Poisson distribution was invented to describe
the the number of random events expected to occur
in a given time interval.  One of the first uses of it 
was to describe the number of Prussian cavalry which would be  
kicked to death by horses in a given year!  The actual numbers
matched the distribution very closely.
}
$p(k) = (\mu^k e^{-\mu})/k!$, where $p(k)$ is the probability of
exactly $k$ counts and $\mu$ is the mean of the distribution.
Also, if we were measuring the number of times that Alice and 
Bob ``simultaneously'' observe a photon in that 10 seconds,
that would be expected to follow a Poisson distribution (with a 
different mean).  Here ``simultaneously'' means that Alice
and Bob both observe photons at times differing by less than some 
preassigned constant $\delta > 0$; e.g., they both observe a 
photon at times differing by less than 1 microsecond. 
But these are not what we are measuring.

What we are measuring is the following.
First we select all the occasions on which Alice and Bob
receive a photon ``simultaneously'' (as defined in the last paragraph).
Then for each such occasion, we observe the value of a ``yes-no''
random variable which takes the value ``yes'' if and only if 
(Alice observes spin $+1$ (relative to her instrument set at $\avec$)
and Bob observes spin $+1$ (relative to his instrument set at $\bvec$))
{\em or} (Alice observes spin $-1$ and Bob also observes $-1$).
Then we calculate the relative frequency of ``yes'' answers
(the number of occurrences of ``yes'' divided by the total
number of simultaneous pairs), a statistic $S$ called the ``sample mean''
(to distinguish it from the usually unknown mean of the probability
distribution from which the random sample is drawn). 
The sample mean $S$ estimates the probability (call it $q$)
of ``yes''.  Routine calculation reveals that when $n$ simultaneous
pairs are observed, the sample mean
has standard deviation $\sqrt{q(1-q)}/\sqrt{n}$
Hence it seems reasonable to estimate the standard deviation of 
the sample mean by%
\footnote{
All of this is standard statistics.
For simplicity, 
I am glossing over some statistical subtleties which are 
unimportant in the present context.  
For example, calculation reveals that the estimator $S(1-S)/n$ of the  variance of the sample mean 
is (surprisingly) not ``unbiased''; to get an unbiased estimator
one replaces  ${S(1-S)}/{n}$ by ${S(1-S)}/(n-1)$.
For large $n$, the difference is negligible.
It is usual to estimate the standard deviation of the
sample mean as the square root of the estimator for the variance even though
this estimator is not unbiased with either estimator
of the variance. 
} 
$$
\frac{\sqrt{S(1-S)}}{\sqrt{n}}
\q.
$$

From this, follows easily an estimate for the correlations   
$C := C(\avec, \bvec) = E(\avec, \bvec)$.%
\footnote{I am following the paper in assuming that $C(\avec, \bvec)
= E(\avec, \bvec)$, where $E(\avec, \bvec)$ denotes the average of 
$C(\avec, \bvec)$ over the plane of $\avec, \bvec$.  The next section
wonders about this assumption.
}
Suppose that we observe
$n$ photon pairs with $n_+$ ``yes'' results and $n_- $ ``no'',
$n_+ + n_- = n$ .  Then the sample mean $S = n_+ / n$, and the 
measured correlation $C = n_+ /n - n_-/n = (2n_+ - n)/n = 2S - 1$.
Hence the estimated standard deviation of $E = C$ is twice the estimated 
standard deviation $\sqrt{S(1-S)/n}$ for $S$.    

We can't apply this directly to the results of the paper because
the value of $n$ (number of photon pairs used to calculate
the sample mean) is not given.  However, we can ask
what value of $n$ would yield the paper's claimed error of
.0118 for $E(\avec_2 , \bvec_3 ) = -.9902 \pm .0118$ (bottom of p.\ 6).
The claimed error [standard deviation] of .0118 for 
$C = E := E(\avec_2 , \bvec_3) $ corresponds
to a standard deviation of .0059 for $S$, so we need to solve the equation
$$
\frac{\sqrt{S(1-S)}}{\sqrt{n}} = .0059
$$
with $S := (C+1)/2 = (E+1)/2 = .0049$.  

The solution is $n \approx 140$, which seems rather small.  
The paper mentions approximately 3000 photon pairs received in 
10 sec.  If this were the true value of $n$, then the claimed
error of .0018 for $E(\avec_2, \bvec_2)$, which scales with $1/\sqrt{n}$,
would be about 5 times smaller.  I wonder if the paper may have
inadvertently {\em overstated} the errors. 
\section{Final comments}
As a mathematician who is largely self-taught in physics,
I am unsure of the correspondence between the physical measurements
described in the paper and the mathematics of the Poincar\'{e} sphere.
Is this well-established physics, or is it a kind of guess,
based on mathematical analogies between complex polarization vectors
in classical electrodynamics and  the two-dimensional complex 
state space describing quantum-mechanical photons?

I am uneasy about 
the paper's justification for its assumption that the average 
over a great circle on the Poincar\'{e} sphere can be confidently replaced 
by an evaluation of the  single correlation $C(\avec, \bvec)$ for 
$\avec, \bvec$ on the circle.
The paper justifies this assumption as follows: (bottom of p.\ 5): 
\begin{quote}
``So far, no experimental evidence against the 
rotational invariance of the singlet state exists.  We therefore 
replace the rotation averaged correlation functions in inequality (9)
with their values measured for one pair of settings (in the given plane).''
\end{quote}
It seems dangerous to assume that something is true 
on the sole grounds that no one has proved it false.  
That risks overlooking potentially important new physics.

My impression is that $C(\avec, \bvec) = -\avec \cdot \bvec$ 
is experimentally well established 
for correlations $C(\avec, \bvec)$ with $\avec$ and $\bvec$  
in the $x$-$z$ plane, i.e., linear polarizations.
I'm not aware of any experiments explicitly validating it for
$\avec, \bvec$ lying in some other plane.
Are there any? If so, it would be helpful if the paper gave 
references.

The results of the paper suggest its confirmation
for the $y$-$z$ plane in that correlations in the $y$-$z$ plane
are used in calculating $S_{NLHV}$ on the left side of inequality (9),
and the measured values of $S_{NLHV}$ are consistent with 
quantum mechanics.  However, the actual measured correlations
$C(\avec, \bvec)$ are not given in the paper, except for
a few special cases at the bottom of p.\ 6.  

Enough data 
to suggestively confirm $C(\avec, \bvec) = - \avec \cdot \bvec$
for the $y$-$z$ plane
was probably gathered in the course of the experiment.  
It would have been helpful had it been presented,
if not in the Nature article (which might have had length constraints),
then in an arXiv report.  These experiments are probably 
hard to do, and print is cheap. 
 
\end{document}